\documentclass[pre,showpacs,amsmath,amssymb,reprint,superscriptaddress, nofootinbib]
{revtex4-1}

\usepackage{graphicx}
\usepackage{amsmath,amssymb}
\usepackage{array}
\usepackage{color}
\usepackage{multirow}
\usepackage{scalerel}

\setcitestyle{round}
\usepackage[dvipsnames]{xcolor}
\usepackage{comment}

\usepackage{cleveref}

\newcommand{\etal}{{\it et al. }}

\newcommand{\nPerp}{{\nabla^{\perp}}}
\newcommand{\rl}{{r/\lambda}}

\newcommand{\etam}{{\eta_{2D}}}
\newcommand{\etaf}{{\eta_{3D}}}

\usepackage{xcolor}

\makeatletter
\newcommand{\globalcolor}[1]{%
  \color{#1}\global\let\default@color\current@color
}
\makeatother

\begin{document}
\def\hrSys{\scalerel*{\includegraphics{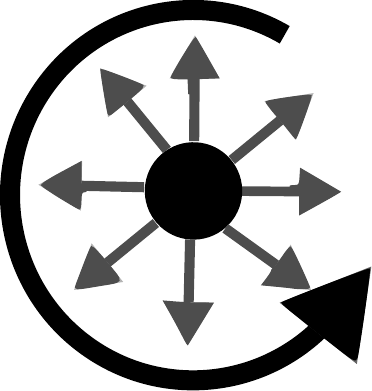}}{S}}
\def\rSys{\scalerel*{\includegraphics{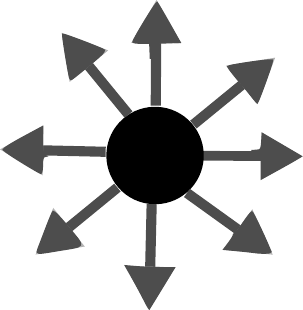}}{S}}
\def\hSys{\scalerel*{\includegraphics{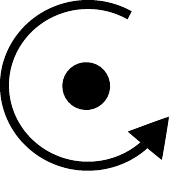}}{S}}
%\pagecolor{black}

\preprint{}

\title{Fast crystallization of rotating membrane proteins}
%\title{Hurricane dynamics in a biomembrane}

% repeat the \author .. \affiliation  etc. as needed
% \email, \thanks, \homepage, \altaffiliation all apply to the current
% author. Explanatory text should go in the []'s, actual e-mail
% address or url should go in the {}'s for \email and \homepage.
% Please use the appropriate macro foreach each type of information

% \affiliation command applies to all authors since the last
% \affiliation command. The \affiliation command should follow the
% other information
% \affiliation can be followed by \email, \homepage, \thanks as well.

\author{Naomi Oppenheimer}
\email{naomiop@gmail.com}
\affiliation{Center for Computational Biology, Flatiron Institute, New York, NY 10010, USA}
\author{David B. Stein}
\affiliation{Center for Computational Biology, Flatiron Institute, New York, NY 10010, USA}
\author{Michael J. Shelley}
\email{mshelley@flatironinstitute.org}
\affiliation{Center for Computational Biology, Flatiron Institute, New York, NY 10010, USA}
\affiliation{Courant Institute, New York University,
New York, NY 10012, USA}

\date{\today}

\begin{abstract}
We examine the interactions between actively rotating proteins moving
in a membrane. Experimental evidence suggests that such rotor
proteins, like the ATP synthases of the inner mitochondrial membrane,
can arrange themselves into lattices. We show that crystallization is possible
through a combination of hydrodynamic and repulsive interactions
between the rotor proteins. In particular, hydrodynamic interactions induce rotational motion of the 
rotor protein assembly that, in the presence of repulsion, drives the system into a hexagonal lattice. The entire crystal rotates with an angular velocity which increases with motor density and decreases with 
lattice diameter --- larger and sparser arrays rotate at a slower pace. 
The rotational interactions allow ensembles of proteins to sample configurations and reach an ordered steady state, which are inaccessible to the quenched nonrotational system. Rotational interactions thus act as a sort of temperature that removes disorder, except that actual thermal diffusion leads to expansion and loss of order. In contrast, the rotational interactions are bounded in space. Hence, once an ordered state is reached, it is maintained at all times. 

\end{abstract}

% insert suggested PACS numbers in braces on next line
% insert suggested keywords - APS authors don't need to do this
%\keywords{}

%\maketitle must follow title, authors, abstract, \pacs, and \keywords
\maketitle
%=================================================================================

Biological membranes serve as barriers between the cell and the outer
environment. Unlike most barriers, biomembranes are fluid --- 
their constituent lipids are free to flow in the plane of the
membrane \cite{singer1972}. There is immense significance to this
fluidity as it enables cell signalling, cell division, the formation
of lipid rafts and more \cite{Phillips2013}. Moreover, the membrane is
not a strictly two dimensional (2D) fluid, at large distances the
fluids outside and inside the cell influence the flow in the plane of
the membrane. Thus, the membrane has mixed dimensionality,
transitioning between two-dimiensional behavior at small distances
to three dimensional (3D) behavior at large distances.  The typical
distance where this transition occurs was first predicted by Saffman
and Delbr\"uck \cite{Saffman1975} and is given by $\lambda =
\eta_{2D}/(2\eta_{3D})$, where $\eta_{2D}$ is the 2D viscosity of the
membrane and $\eta_{3D}$ is the 3D viscosity of the outer fluid. For
biological membranes this length is around a micron, much larger than
the typical size of a lipid ($\sim 1$ nm) or a motor ($\sim 10$ nm).

\begin{figure}[h]
%\vspace{0.2cm} 
\includegraphics[width=0.4\textwidth]{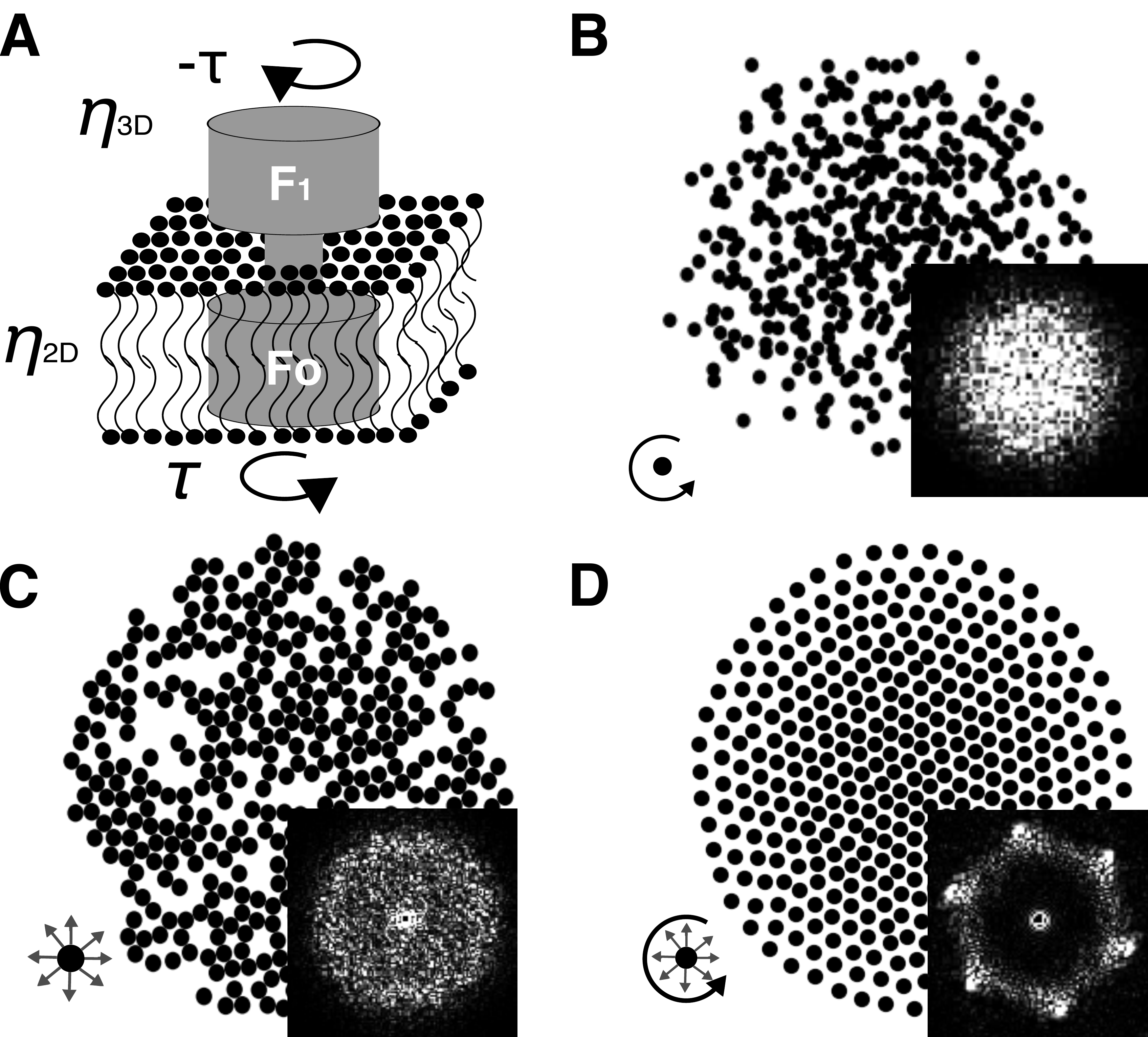} 
\caption{(a) A representation of a membrane protein as two
  counter-rotating disks, each of radius $a$, one in the plane of the
  membrane, the other in the outer fluid, with torque strengths
  $\tau$ and $-\tau$ respectively. (b-d) Snapshots of three
  simulations, starting with the same initial conditions and taken at the same time-point, of (b) rotor proteins with only rotation and no repulsion (c) only repulsion and no rotation, and (d) rotor proteins with both rotation and repulsion. Inset of each
  figure is the structure factor showing hexatic ordering for system
  (d) but no distinct ordered structure for systems (b) and (c). }
\label{membraneCartoon}
\end{figure}

A biological membrane is not a homogeneous fluid; it includes around
thirty percent transmembrane proteins, making the membrane a quasi-2D
suspension \cite{Oppenheimer2009}. These proteins play various roles
and exhibit myriad behaviors. Some proteins passively diffuse in
the plane of the membranes \cite{Levine2004}, while others are actively driven
\cite{Gov2004,Hosaka2017}. In this Letter, we examine the flows and patterns created by one very important transmembrane protein --- ATP synthase, a protein that creates the cellular
fuel ATP (adenosine triphosphate) by phosphorylation of ADP (adenosine
diphosphate), and which rotates as part of its function
\cite{Davies2012}. ATP synthase has two counter-rotating
structures, the ${\rm F_O}$ subunit which sits primarily in the
membrane, and the ${\rm F_1}$ subunit which protrudes into the
outer fluid; see Fig.~\ref{membraneCartoon}a. The rotation rates are
proportional to the flux of protons through the protein (and thence across
the membrane) and range between $10^{2-3}$~Hz \cite{Ueno2005,Martin2014}. 
Such {\it rotor proteins} are a self-driven active-matter system \cite{Nguyen2014,Yeo2015,Soni2018}. Since no external torque acts upon the protein its induced flow resembles a torque-dipole to a leading order \cite{Lenz2004}.

%ensemble, 
Electron cryomicroscopy of ATP synthase in lipid vesicles shows that
rotor proteins can form lattices in the flat regions of the vesicle
\cite{Jiko2015}. By studying infinite systems of model rotor proteins,
it was suggested earlier by Lenz \etal \cite{Lenz2003,Lenz2004} that a
hexagonal lattice is a neutrally stable state. For finite systems we
show that adding repulsive interactions leads to a lattice state even
when starting from random initial conditions.
Figure~\ref{membraneCartoon}b--d shows the central result of this work --- snapshots of three different simulations of rotor protein assemblies,
which interact via different combinations of repulsion and rotation: panel b (marked \hSys), only hydrodynamic interactions; panel c (marked \rSys), only repulsive
interactions; and panel d (marked \hrSys), both hydrodynamic and repulsive interactions. The insets show the structure factor
of each system, given by $S({\bf q}) =  \sum_{i,j} e^{- i {\bf q} \cdot ({\bf r}_i - {\bf r}_j)} $ \cite{Chaikin}. The six distinct peaks in the structure factor of
System \hrSys~indicate the presence of global hexagonal order, versus the lack of
distinct global features in the structure factors of Systems \hSys ~and \rSys.

\smallskip
{\bf The hydrodynamics model and its properties:} To understand the dynamics of a rotor protein assembly, 
let us describe a simpler starting point --- a membrane rotor, modeled as a single disk of radius $a$ rotating %with an angular velocity $\omega$ 
due to an external torque $\tau$, and sitting in a flat membrane whose
velocity is ${\bf v}$. We assume membrane incompressibility ($\nabla \cdot {\bf v}= 0$), and negligible inertia (small Reynolds number). Under these assumptions, momentum conservation in the membrane reads,
\begin{equation}
0 = \etam \nabla^2 {\bf v} + \etaf \left [ \frac{\partial {\bf u}}{\partial z}\right]_{z=0} + \tau \nPerp \delta({\bf r}),
\end{equation}
where ${\bf  u}$ is the 3D flow in the outer fluids, and $\nPerp =
(-\partial/\partial y,\partial/\partial x)$. The second term on the
right hand side is the jump in shear stress from the outer fluids, and the
third term is the force due to a point torque. There is no pressure
contribution for purely rotational motion, or a superposition of such
flows. The outer fluids obey the 3D Stokes equations with the boundary condition 
${\bf  u}^\pm|_{z=0} = {\bf v}$. It is easy to solve this coupled system
for the 2D stream-function $\Psi$ of the velocity ${\bf v}$ using
a 2D Fourier transform ($\tilde{F}({\bf q}) = \int\int F({\bf r})
e^{i {\bf q} \cdot {\bf r}} d^2r$), giving:
\begin{equation}
  \tilde{v}(q) = \frac{\tau}{\etam} \nPerp \tilde{\Psi} \ \ \ ;
  \ \ \  \tilde{\Psi} = \frac{1}{q (q+\lambda^{-1})}.
\end{equation}
In real space $\Psi({\bf r}) = 1/4 (H_0(\rl) - Y_0(\rl))$, 
where $H_0$ and $Y_0$ denote the order zero Struve function and 
Bessel function of the second kind, respectively.

To derive the stream function for a rotor protein, a second counter-rotating
disk is placed a distance $l$ away in the outer fluid,
such that the total torque on the protein is zero (see \Cref{membraneCartoon}A). A similar,
albeit lengthier derivation, assuming $r\gg l$ (see Supplementary Information
and Ref.~\cite{Lenz2003}), gives an almost identical result for the
stream function but with one additional term,
\begin{eqnarray}
{\bf v}({\bf r}) &=& \Gamma \nPerp \Psi \nonumber \\ 
\Psi ({\bf r}) &=&  \frac{1}{2 \pi} \left[\frac{\lambda}{r}+ \frac{\pi}{2} \left( Y_0\left(\frac{r}{\lambda}\right) - H_0\left(\frac{r}{\lambda}\right)\right)\right],
\label{streamMotor}
\end{eqnarray}
where $\Gamma=\frac{2\tau l }{\etam \lambda}$. 
We can now consider the scaling of Eq.~(\ref{streamMotor})
in the various limits. For small distances compared to the
Saffman-Delbr\"uck (SD) length, $r\ll \lambda$, the stream function
satisfies $\Psi\sim 1/r$. Therefore $v
\sim \nPerp \Psi \sim 1/r^2$. Qualitatively, this can be understood as
follows: in a 2D fluid, a Stokeslet (the flow due to a point force $\delta({\bf
  r})$) scales as $\log r$; a rotlet (the flow due to a point torque $\nPerp \delta({\bf r})$) thus scales as  $v \sim 1/r$. The flow due to a torque dipole scales as yet
another derivative, leading to $v\sim 1/r^2$. In the opposite
limit, $r \gg \lambda$, the 3D fluid dominates. The flow due
to a point force at large distances scales as $1/r$, we would expect
the flow due to a rotating protein to scale as $1/r^3$, but this term
cancels due to symmetry at the $z=0$ plane, and so we have $v
\sim 1/r^4$.

For more than one rotor protein, the stream function can be
generalised to the Hamiltonian $\mathcal{H} = \sum_{i\neq j} \Gamma_i \Gamma_j \Psi(|r_i - r_j|)$, where $\Gamma_i = \frac{2\tau_i l
}{\etam \lambda}$ is the strength of the $i$th torque dipole. The
velocity is given by ${\bf v}_i = (1/\Gamma_i) \nabla_i^{\perp}
\mathcal{H}$ \cite{Lushi2015}. Note that $\mathcal{H}$ is a Hamiltonian in the 2D coordinates, $x$ and $y$,
(unlike in classical mechanics where it is a function of
position and conjugate momentum), and so phase-space
corresponds to the positions of the rotors. From Noether's theorem
\cite{Noether1918}, symmetries of the Hamiltonian correspond to
conservation laws. In our case there is conservation of the
Hamiltonian itself, and of the first- and second-moments (from time,
translational and rotational invariance respectively). 

We take all the ``circulations", $\Gamma_i$, to be equal, $\Gamma_i = \Gamma$, and
hence same-signed\footnote{For ATP
synthase, angular velocity of the proteins is a function of proton
concentration. Since protons are seven orders of magnitude smaller
than the proteins, we can assume that
proton concentration equilibrates on a much faster time-scale.}.  
Conservation of the second-moment
then simplifies to $\mu_2 = \sum_{i,j}|r_i - r_j|^2 = {\rm Const}$. 
As a result of this invariance the distance between rotor proteins cannot diverge to infinity, and by invariance of $\mathcal{H}$ the distance cannot collapse to zero. In general, rotor protein ensembles stay bounded in an area not much different than their initial area, and maintain a finite distance between each other.
A crude upper bound for the radius of the rotating system can be derived as follows. For a random initial condition, the second moment is proportional to the initial area and the number of particles squared, i.e. $\mu\sim r_0^2N^2$. The maximal radius is bounded by a configuration in which all particles but one are at the origin. The remaining particle will then have the maximal possible distance from the origin, given by $r_{\rm max} \leq \sqrt{\mu_2/ N} \propto r_0 \sqrt{N}$. This bound can be improved by incorporating conservation of the Hamiltonian, see Supplementary Information.

Surprisingly, the dynamics of such a system is very similar to that of point singularities in 2D such as the ideal vortices of a 2D Euler fluid, or those of the quasi-geostrophic (QG) equations which arise in modeling atmospheric flows \cite{Falkovich2009, Conti2017}. In all such systems two singularities will orbit around each other, and for four or more the system becomes non-integrable and the dynamics can be chaotic \cite{Aref1982} (see SI for a video of typical rotor protein dynamics). Indeed, the near-field interactions for rotor proteins are the same as for QG vortices.

In what follows we consider two aspects of membrane proteins that are not captured by pure hydrodynamic interactions. First, membrane proteins are physical objects of finite size with possible interactions with other proteins. Second, due to their small size ($\sim10$ nm) thermal noise might play a significant role in their dynamics. We show that adding any type of repulsion between the rotor proteins can result in the formation of crystals; sufficiently high temperature can destroy that order.

\smallskip
{\bf Repulsive interactions drive crystalization}. To start, we follow
\cite{Lenz2003} who noted that for a system with only hydrodynamic interactions (System \hSys), an infinite hexagonal array of rotor
proteins is a steady-state of the system, for any lattice scale $d$. Given the system's Hamiltonian structure, this array is, at most, neutrally stable. This means that if the system does not start in an ordered array it will not reach that state. For more details see the SI. 
Adding repulsive interactions (System \hrSys) can turn this neutrally stable
fixed point into a stable one. For example, a linear perturbation, $\delta r(t)$, on a single rotor protein must have the form,   $d\delta {\bf r}/dt = -\alpha
\delta {\bf r}_{\perp} - \beta \delta {\bf r}$, where $\alpha$ ($\beta$) is coming from the rotational (repulsive) interaction. The eigenvalues of
this system are $\lambda = \pm i \alpha - \beta$. The additional
negative component is a necessary condition for linear stability of the hexagonal configuration. From simple
symmetry arguments it is clear that an infinite or a confined system
with long ranged repulsive interactions must form a hexagonal
lattice, also known as a Wigner crystal \cite{Wigner1934}.

However, this is not the case for a finite system with 
only short-ranged repulsive interactions, as can be seen from
Fig.~\ref{membraneCartoon}c. Surprisingly, experiments by
Jiko \etal \cite{Jiko2015} still show an ordered formation even for finite ensembles of proteins.
We consequently examined the dynamics of a finite, unconfined system of rotor proteins with short-ranged repulsion with an interaction distance $r_s$, using either exponential (e.g. electrostatically screened interactions) $U = U_0 e^{-r/r_s}$, or soft, harmonic-like
repulsion of the form $U = U_0 (r-r_s)^2$ up to an inter-particle distance of $r_s$, and zero otherwise. 

Starting from random initial conditions and letting the system
evolve, we discovered that rotation promotes rapid organization into
crystals even in cases where the repulsion alone is not sufficient to
induce the formation of a lattice, see Figs.~\ref{membraneCartoon}b
versus~\ref{membraneCartoon}d. In system \rSys, a particle moves until it no longer feels its neighbours. At low concentrations, the system is quenched in a disordered state. Adding rotation to the particles, system \hrSys, stirs and reshuffles them. Initially the amount of interactions increases, until a steady state configuration is reached in which particles are equally distanced. The whole lattice then rotates as a rigid body. 

\begin{figure}[t]
\vspace{0.2cm} 
\includegraphics[width=0.38\textwidth]{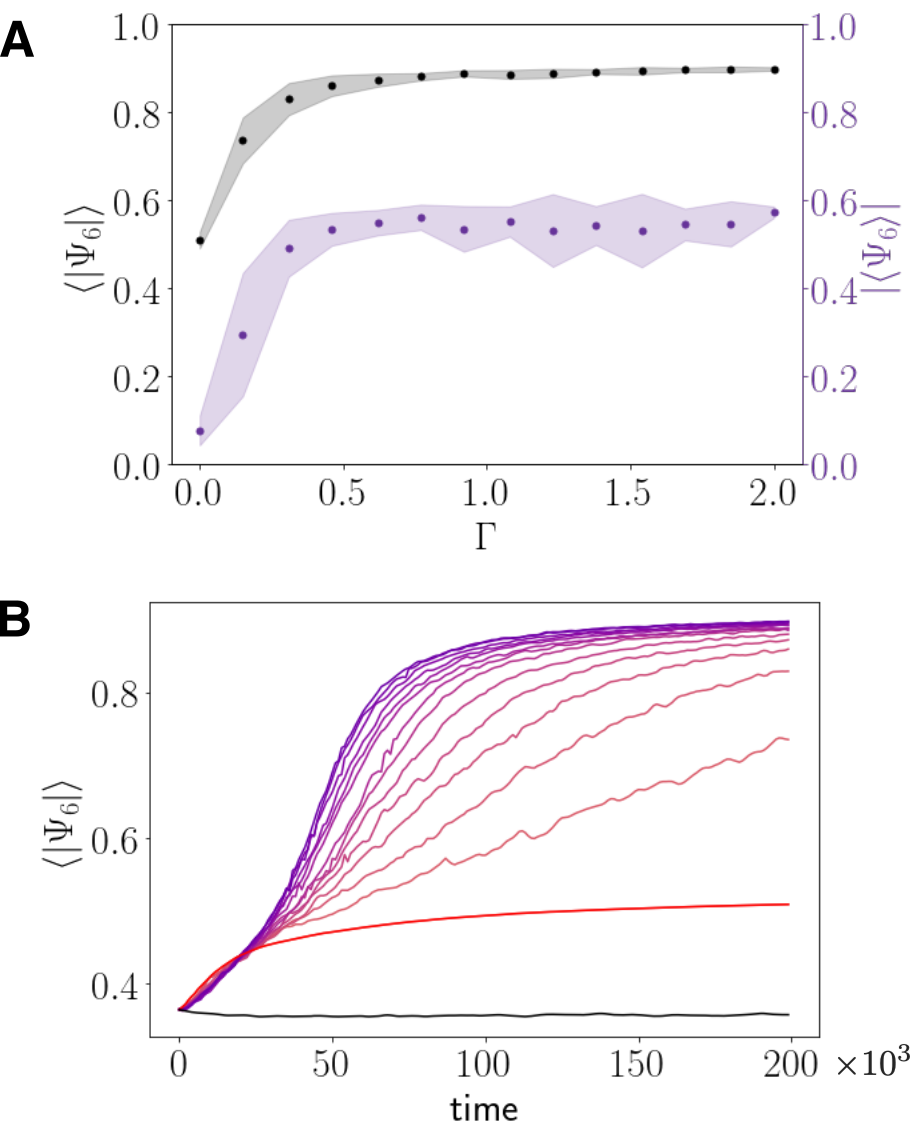} 
\caption{(a) The average local, $\langle|\Psi_6|\rangle$ (black) and global $|\langle \Psi_6\rangle|$ (purple) bond order parameters as a function of circulation, $\Gamma$, calculated after $2\cdot 10^5$ time steps. Each point corresponds to an average over 14 random initial configurations in a system of 400 motors with soft, spring-like repulsion, with a final area fraction of 0.8. The shaded regions are the standard deviation.  
(b) The local order parameter as a function of time. Hue increases with increasing circulation. The black curve corresponds to the case of no repulsion between particles, the red curve corresponds to the case of no rotation.}
\label{orderFigure}
\end{figure}

{\bf Increased rotor activity yields faster ordering}. To measure order in the system we look at the structure factor, $S({\bf q})$, and at the two-dimensional bond-orientational order parameter, $\Psi_6^j = (1/n_j) \sum_i e^{i 6\theta_{ij}}$. $\Psi_6^j$ measures the orientation and degree of hexagonal order around particle $j$ \cite{Nelson2002}, the sum is over the nearest neighbours of particle $j$ as found from Delaunay triangulation \cite{Delaunay1934}, $n_j$ is the number of nearest neighbours, and $\theta_{ij}$ is the angle between the bond connecting particles $i$ and $j$ and the $x$ axis (which is an arbitrary reference). Let $\langle\cdot\rangle$ define averages over all ensemble particles. We define the average local and global order parameters to be $\langle|\Psi_6|\rangle$ and $|\langle\Psi_6\rangle|$, respectively. The global order parameter is close to zero for all initial random configurations, and is one for a perfect infinite hexagonal lattice. 
As shown in Fig.~\ref{orderFigure}, purely repulsive interactions lead to slight increase in the order parameter, but do not result in an ordered array. Increasing the circulation, $\Gamma$, increases the order in the system, saturating the average local bond orientational parameter at a value of around 0.9 for large circulations. Repulsive interactions promote the formation of an ordered state but are unnecessary for maintaining order: once the system has reached the state shown in Fig.~\ref{membraneCartoon}d, turning off the repulsive interactions does not destroy the order. This is a manifestation of the neutral stability of the purely hydrodynamic system, \hSys. 
 
\begin{figure}[h]
\vspace{0.2cm} 
\includegraphics[width=0.45\textwidth]{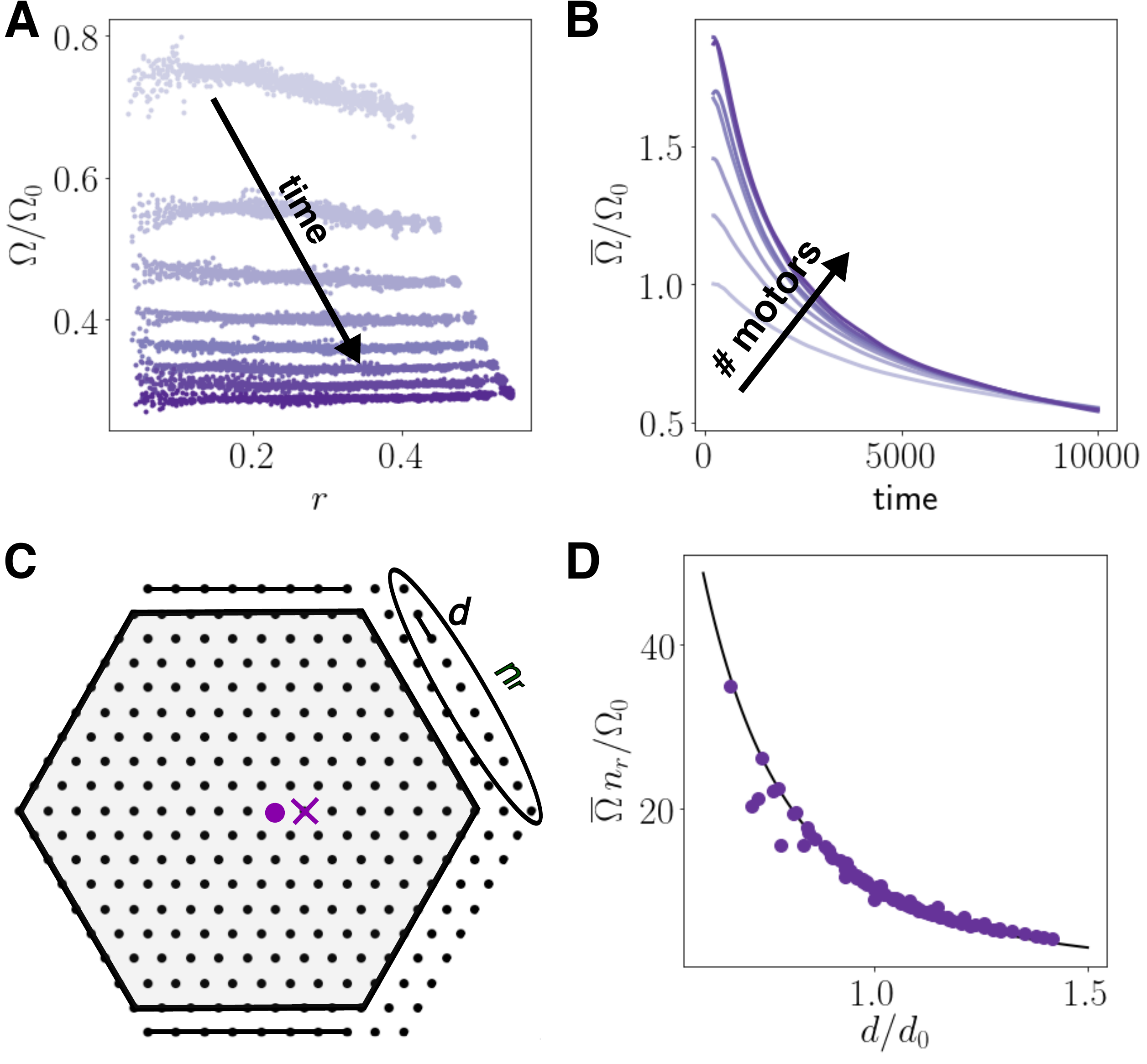} 
\caption{(a) Angular velocity, $\Omega$, as a function of radius, increased hue corresponds to the progression of time. Density decreases with time, resulting in lower angular velocity. At long times the lattice rotates as a rigid body. (b) Increase in hue corresponds to increasing number of rotors, going from 100 to 1200 rotors. (c) Assuming rigid body rotation and a perfect, finite, hexagonal lattice, whose center is marked with an $x$, the angular velocity of the point marked with a dot is equal to that of the entire crystal. From symmetry, any motor in the grey hexagon or the black lines will not contribute. The only contribution comes from the four rows of proteins at the edges. (d) scaling the average angular velocity according to Eq.~\ref{omega}, all results fit on a single curve. 
Angular velocity in all subfigures are normalized by $\Omega_0 = \Omega(t=0)$ for 200 rotors.}
\label{velocityFigure}
\end{figure}

{\bf Estimating the lattice rotation rate}. Once formed, the crystals are not stationary in space, but rotate around their center of mass with an angular velocity $\Omega(r)$. 
Figure~\ref{velocityFigure} shows the angular velocity as a function of radius for a finite system of particles interacting via Eq.~\ref{streamMotor} and with exponential repulsive interactions. 
We note two things about the angular velocity.  First, it decreases with time  (Fig.~\ref{velocityFigure}a). This decrease is a result of the collective expansion driven by the exponentially repulsive interactions. Second, changing the number of rotor proteins, but keeping the initial area fixed results in an increase in $\Omega$, as can be seen in Fig.~\ref{velocityFigure}b. In order to explain these features, we construct a simplified model based on two observations from  Figures \ref{membraneCartoon}d and \ref{velocityFigure}a: 1) the lattice rotates as a nearly rigid body, and 2) it forms a nearly perfect lattice (with spacing that only weakly depends on the radius). We hence assume rigid body rotation of a perfect lattice with spacing $d$ and angular velocity $\bar{\Omega}$. 
With these simplifications at hand, consider the following argument: for an infinite system the velocity must be zero from symmetry. 
To see this, consider the central particle in Fig.~\ref{velocityFigure}c (marked with a cross), each two opposing particles surrounding it will create opposite flows resulting in zero net velocity. 
Now note that for an infinite system all particles are identical, therefore the velocity of each particle in an infinite lattice is zero. For a finite system, it is thus the perimeter which dictates the angular velocity of the lattice, namely the number of proteins on the edge, $n_r$, and the distance between them, $d$, see Fig.~\ref{velocityFigure}c. 
Specifically, in the limit of small distances, $ R \ll \lambda$, the velocity of the $k$th protein in complex notation, $z = x+ i y$, is given by
\begin{equation}
\frac{dz_k}{dt} = i \sum_j \frac{z_k - z_j}{|z_k - z_j|^3},
\end{equation}
where distance is normalized by $\lambda$, and time by $2\pi \lambda^2/\Gamma$.
Assuming the angular velocity is the same for all proteins, we can calculate the induced angular velocity of a single protein, for example, the particle just left of the center, $z_k = - D e^{i \Omega t}$ (marked with a disk in Fig.~\ref{velocityFigure}c), where $D = d/\lambda$. 
The only components of the sum which do not cancel are the four rows at the edges (see Fig.~\ref{velocityFigure}). For a large number of particles, $n_r\rightarrow \infty$, the sum simplifies to,
\begin{equation}
 \Omega  \xrightarrow{N\rightarrow \infty}  \frac{4}{D^3}  \sum_{j=0}^{n_r} \frac{\cos\left( \frac{\pi n}{3n_r}\right)}{n_r^2} =  \frac{6 \sqrt{3} }{\pi D^3  n_r}
\label{omega}
\end{equation}
Similarly, in the opposite limit, of proteins separated by distances larger than the SD length, $\lambda$, we get $\Omega \propto 1/(D^3 n_r^3)$,
(see supplementary information for more details). 
A more precise calculation, accounting for density and angular velocity variations as a function of radius, will lead to correction of the prefactor in Eq.~\ref{omega}, but the scaling will remain. After scaling the angular velocities at different times and for different system sizes according to Eq.~\ref{omega}, all results fall on a single curve, see Fig.~\ref{velocityFigure}d. 

\begin{figure}[h]
\includegraphics[width=0.45\textwidth]{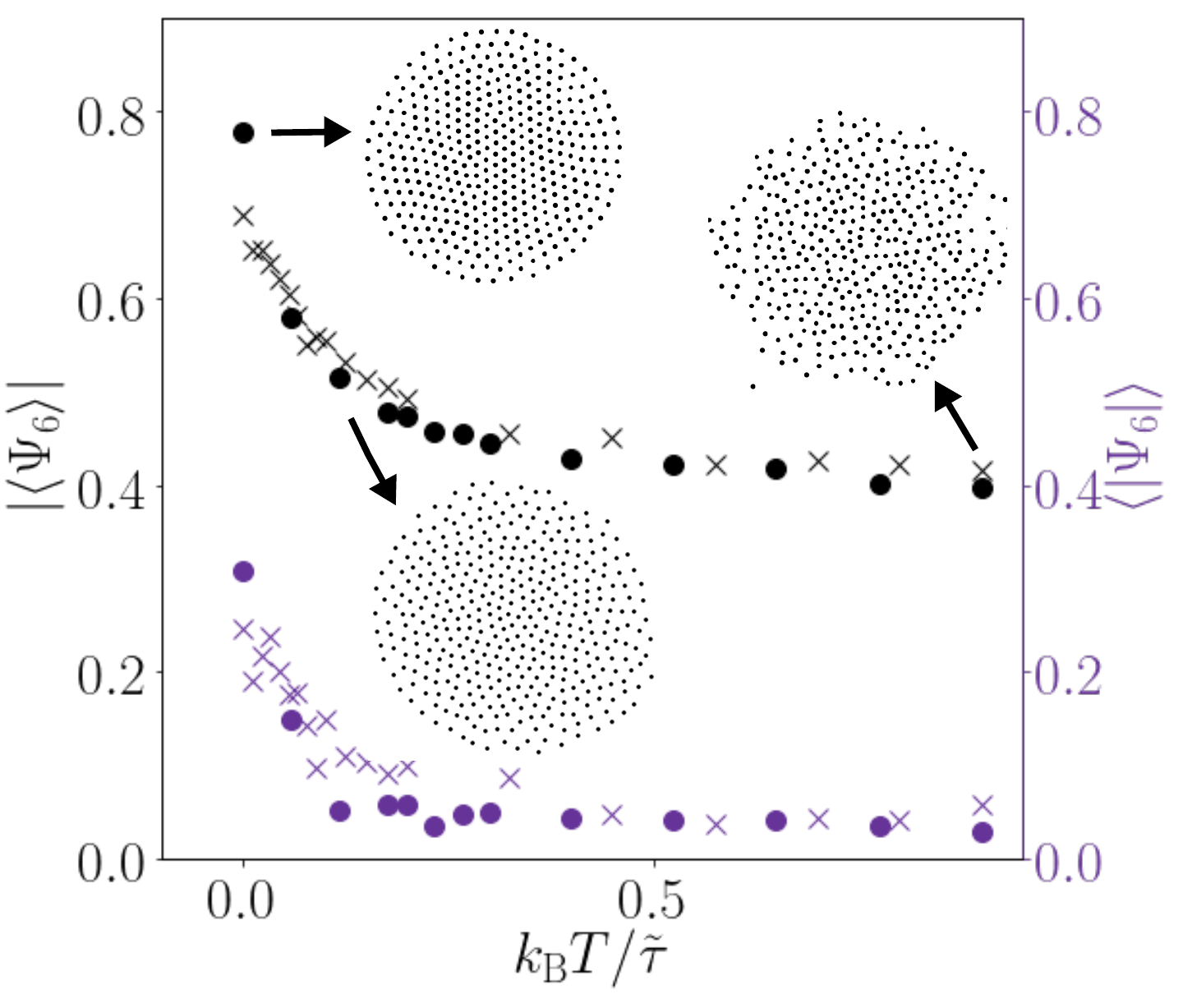} 
\caption{Temperature effect on the global (in purple) and local (in black) bond order parameters in a system of 200 (crosses) and 400 (disks) particles. 
Results are given for an average over nine random initial configurations. Snapshots of the system at three different temperatures, taken at the same time, is shown.}
\label{temperatureFigure}
\end{figure}

{\bf Effect of thermal noise}. We have neglected thermal fluctuations so far but proteins are small ($\sim 10$nm) and prone to thermal forces. Thermal fluctuations must be added carefully when hydrodynamic interactions are included, as the thermal motion of one protein will effect that of another via the fluid in which they are immersed. To account for this, we use the positive definite analogue of the Rotne-Prager mobility tensor for a membrane, given in Ref.~\cite{Sokolov2018}.
Figure~\ref{temperatureFigure} presents the results of increasing temperature on the global and local bond orientational order parameter for two different systems (of 200 and of 400 rotor proteins). At low temperatures the systems are ordered, transitioning into disordered ones at about $k_{\rm B} T/\tilde{\tau} \sim 0.3$, where $\tilde{\tau} = \etam \Gamma$ is the torque-dipole strength acting on a rotor protein.

To estimate the biologically relevant regime, we take the P\'eclet number to be ${\rm Pe} = v a/ D$, where $v$ is the hydrodynamic advective velocity of Eq.~\ref{streamMotor} summed over all particles in the ensemble, and $D$ is the diffusion coefficient of a rotor protein. Using measurements in the literature for membrane viscosity, protein rotation rate and size \cite{Kinosita1998, Levitt2009,Ueno2005}, we estimate the P\'eclet number for ATP synthase to be around ${\rm Pe} \sim 2$. In our simulations, this corresponds to a normalized energy of about $k_{\rm B} T/\tilde{\tau} \sim 0.1$, which is below the transition point. This implies that biological systems should be in a relatively ordered state, but that thermal noise is also not negligible (see the middle snapshot in Fig.~\ref{temperatureFigure}). Note however that this estimate does not incorporate the repulsive interactions, which would have their
own P\'eclet number, and which we expect to further stabilize the system to thermal forces.
Even in the absence of complete order, strong hydrodynamic interactions due to the rotation of the proteins may aid in mixing or in transporting other materials in the membrane.

{\bf Discussion}. To conclude, a system of rotor proteins self organizes into a hexagonal lattice through a combination of hydrodynamic and repulsive interactions. 
At low rotor protein concentrations, with only repulsion, the system (\rSys) is quenched in a disordered state. Adding rotation (\hrSys), and hence hydrodynamic interactions, shuffles the positions of the particles, driving them to an ordered steady state. In this sense rotation is similar to temperature, allowing the system to sample phase-space. Unlike temperature, the rotational interactions, in the absence of repulsive interactions, preserve the second moment as well as the Hamiltonian. The rotating system is then self-confined, whereas with temperature, an initial configuration will spread to infinity as the square root of time.

Throughout this work, we assumed the membrane to be completely flat. In the case of huge vesicles such as those used in \cite{Jiko2015}, and for small thermal fluctuations, this assumption is reasonable. In a biological setting, however, ATP synthase is abundant in the highly curved mitochondrion. It was found that along the mitochondria ridges, ATP synthase dimers form a zigzag topology \cite{Blum2019}, and that there is coupling between the membrane curvature and the activity of ATP synthase \cite{Almendro-Vedia2017}. In the future, we plan to investigate the hydrodynamic coupling between rotation and curvature. 

{\bf Acknowledgment}.
We thank Matan Yah Ben Zion, Enkeleida Lushi, and Florencio Balboa Usabiaga for useful discussions.

\bibliography{library} 

%merlin.mbs apsrev4-1.bst 2010-07-25 4.21a (PWD, AO, DPC) hacked
%Control: key (0)
%Control: author (72) initials jnrlst
%Control: editor formatted (1) identically to author
%Control: production of article title (-1) disabled
%Control: page (0) single
%Control: year (1) truncated
%Control: production of eprint (0) enabled
\begin{thebibliography}{30}%
\makeatletter
\providecommand \@ifxundefined [1]{%
 \@ifx{#1\undefined}
}%
\providecommand \@ifnum [1]{%
 \ifnum #1\expandafter \@firstoftwo
 \else \expandafter \@secondoftwo
 \fi
}%
\providecommand \@ifx [1]{%
 \ifx #1\expandafter \@firstoftwo
 \else \expandafter \@secondoftwo
 \fi
}%
\providecommand \natexlab [1]{#1}%
\providecommand \enquote  [1]{``#1''}%
\providecommand \bibnamefont  [1]{#1}%
\providecommand \bibfnamefont [1]{#1}%
\providecommand \citenamefont [1]{#1}%
\providecommand \href@noop [0]{\@secondoftwo}%
\providecommand \href [0]{\begingroup \@sanitize@url \@href}%
\providecommand \@href[1]{\@@startlink{#1}\@@href}%
\providecommand \@@href[1]{\endgroup#1\@@endlink}%
\providecommand \@sanitize@url [0]{\catcode `\\12\catcode `\$12\catcode
  `\&12\catcode `\#12\catcode `\^12\catcode `\_12\catcode `\%12\relax}%
\providecommand \@@startlink[1]{}%
\providecommand \@@endlink[0]{}%
\providecommand \url  [0]{\begingroup\@sanitize@url \@url }%
\providecommand \@url [1]{\endgroup\@href {#1}{\urlprefix }}%
\providecommand \urlprefix  [0]{URL }%
\providecommand \Eprint [0]{\href }%
\providecommand \doibase [0]{http://dx.doi.org/}%
\providecommand \selectlanguage [0]{\@gobble}%
\providecommand \bibinfo  [0]{\@secondoftwo}%
\providecommand \bibfield  [0]{\@secondoftwo}%
\providecommand \translation [1]{[#1]}%
\providecommand \BibitemOpen [0]{}%
\providecommand \bibitemStop [0]{}%
\providecommand \bibitemNoStop [0]{.\EOS\space}%
\providecommand \EOS [0]{\spacefactor3000\relax}%
\providecommand \BibitemShut  [1]{\csname bibitem#1\endcsname}%
\let\auto@bib@innerbib\@empty
%</preamble>
\bibitem [{\citenamefont {Singer}\ and\ \citenamefont
  {Nicolson}(1972)}]{singer1972}%
  \BibitemOpen
  \bibfield  {author} {\bibinfo {author} {\bibfnamefont {S.~J.}\ \bibnamefont
  {Singer}}\ and\ \bibinfo {author} {\bibfnamefont {G.~L.}\ \bibnamefont
  {Nicolson}},\ }\href {\doibase 10.1126/science.175.4023.720} {\bibfield
  {journal} {\bibinfo  {journal} {Science}\ }\textbf {\bibinfo {volume}
  {175}},\ \bibinfo {pages} {720} (\bibinfo {year} {1972})}\BibitemShut
  {NoStop}%
\bibitem [{\citenamefont {Phillips}(2013)}]{Phillips2013}%
  \BibitemOpen
  \bibfield  {author} {\bibinfo {author} {\bibfnamefont {R.}~\bibnamefont
  {Phillips}},\ }\href@noop {} {\emph {\bibinfo {title} {Physical Biology of
  the Cell}}}\ (\bibinfo  {publisher} {Garland Science},\ \bibinfo {year}
  {2013})\BibitemShut {NoStop}%
\bibitem [{\citenamefont {Saffman}\ and\ \citenamefont
  {Delbr{\"{u}}ck}(1975)}]{Saffman1975}%
  \BibitemOpen
  \bibfield  {author} {\bibinfo {author} {\bibfnamefont {P.~G.}\ \bibnamefont
  {Saffman}}\ and\ \bibinfo {author} {\bibfnamefont {M.}~\bibnamefont
  {Delbr{\"{u}}ck}},\ }\href {\doibase 10.1073/pnas.72.8.3111} {\bibfield
  {journal} {\bibinfo  {journal} {Proc. Nat. Acad. Sci.}\ }\textbf {\bibinfo
  {volume} {72}},\ \bibinfo {pages} {3111} (\bibinfo {year}
  {1975})}\BibitemShut {NoStop}%
\bibitem [{\citenamefont {Oppenheimer}\ and\ \citenamefont
  {Diamant}(2009)}]{Oppenheimer2009}%
  \BibitemOpen
  \bibfield  {author} {\bibinfo {author} {\bibfnamefont {N.}~\bibnamefont
  {Oppenheimer}}\ and\ \bibinfo {author} {\bibfnamefont {H.}~\bibnamefont
  {Diamant}},\ }\href {\doibase 10.1016/j.bpj.2009.01.020} {\bibfield
  {journal} {\bibinfo  {journal} {Biophys. J.}\ }\textbf {\bibinfo {volume}
  {96}},\ \bibinfo {pages} {3041} (\bibinfo {year} {2009})},\ \Eprint
  {http://arxiv.org/abs/0809.4163} {arXiv:0809.4163} \BibitemShut {NoStop}%
\bibitem [{\citenamefont {Levine}\ \emph {et~al.}(2004)\citenamefont {Levine},
  \citenamefont {Liverpool},\ and\ \citenamefont {MacKintosh}}]{Levine2004}%
  \BibitemOpen
  \bibfield  {author} {\bibinfo {author} {\bibfnamefont {A.~J.}\ \bibnamefont
  {Levine}}, \bibinfo {author} {\bibfnamefont {T.~B.}\ \bibnamefont
  {Liverpool}}, \ and\ \bibinfo {author} {\bibfnamefont {F.~C.}\ \bibnamefont
  {MacKintosh}},\ }\href {\doibase 10.1103/PhysRevE.69.021503} {\bibfield
  {journal} {\bibinfo  {journal} {Phys. Rev. E}\ }\textbf {\bibinfo {volume}
  {69}},\ \bibinfo {pages} {021503} (\bibinfo {year} {2004})}\BibitemShut
  {NoStop}%
\bibitem [{\citenamefont {Gov}(2004)}]{Gov2004}%
  \BibitemOpen
  \bibfield  {author} {\bibinfo {author} {\bibfnamefont {N.}~\bibnamefont
  {Gov}},\ }\href {\doibase 10.1103/PhysRevLett.93.268104} {\bibfield
  {journal} {\bibinfo  {journal} {Phys. Rev. Lett.}\ }\textbf {\bibinfo
  {volume} {93}},\ \bibinfo {pages} {268104} (\bibinfo {year}
  {2004})}\BibitemShut {NoStop}%
\bibitem [{\citenamefont {Hosaka}\ \emph {et~al.}(2017)\citenamefont {Hosaka},
  \citenamefont {Yasuda}, \citenamefont {Okamoto},\ and\ \citenamefont
  {Komura}}]{Hosaka2017}%
  \BibitemOpen
  \bibfield  {author} {\bibinfo {author} {\bibfnamefont {Y.}~\bibnamefont
  {Hosaka}}, \bibinfo {author} {\bibfnamefont {K.}~\bibnamefont {Yasuda}},
  \bibinfo {author} {\bibfnamefont {R.}~\bibnamefont {Okamoto}}, \ and\
  \bibinfo {author} {\bibfnamefont {S.}~\bibnamefont {Komura}},\ }\href
  {\doibase 10.1103/PhysRevE.95.052407} {\bibfield  {journal} {\bibinfo
  {journal} {Phys. Rev. E}\ }\textbf {\bibinfo {volume} {95}},\ \bibinfo
  {pages} {052407} (\bibinfo {year} {2017})}\BibitemShut {NoStop}%
\bibitem [{\citenamefont {Davies}\ \emph {et~al.}(2012)\citenamefont {Davies},
  \citenamefont {Anselmi}, \citenamefont {Wittig}, \citenamefont
  {Faraldo-Gómez},\ and\ \citenamefont {Kühlbrandt}}]{Davies2012}%
  \BibitemOpen
  \bibfield  {author} {\bibinfo {author} {\bibfnamefont {K.~M.}\ \bibnamefont
  {Davies}}, \bibinfo {author} {\bibfnamefont {C.}~\bibnamefont {Anselmi}},
  \bibinfo {author} {\bibfnamefont {I.}~\bibnamefont {Wittig}}, \bibinfo
  {author} {\bibfnamefont {J.~D.}\ \bibnamefont {Faraldo-Gómez}}, \ and\
  \bibinfo {author} {\bibfnamefont {W.}~\bibnamefont {Kühlbrandt}},\
  }\href@noop {} {\bibfield  {journal} {\bibinfo  {journal} {Proc. Natl. Acad.
  Sci.}\ }\textbf {\bibinfo {volume} {109}},\ \bibinfo {pages} {13602}
  (\bibinfo {year} {2012})}\BibitemShut {NoStop}%
\bibitem [{\citenamefont {Ueno}\ \emph {et~al.}(2005)\citenamefont {Ueno},
  \citenamefont {Suzuki}, \citenamefont {Kinosita},\ and\ \citenamefont
  {Yoshida}}]{Ueno2005}%
  \BibitemOpen
  \bibfield  {author} {\bibinfo {author} {\bibfnamefont {H.}~\bibnamefont
  {Ueno}}, \bibinfo {author} {\bibfnamefont {T.}~\bibnamefont {Suzuki}},
  \bibinfo {author} {\bibfnamefont {K.}~\bibnamefont {Kinosita}}, \ and\
  \bibinfo {author} {\bibfnamefont {M.}~\bibnamefont {Yoshida}},\ }\href
  {\doibase 10.1073/pnas.0407857102} {\bibfield  {journal} {\bibinfo  {journal}
  {Proc. Natl. Acad. Sci.}\ }\textbf {\bibinfo {volume} {102}},\ \bibinfo
  {pages} {1333} (\bibinfo {year} {2005})}\BibitemShut {NoStop}%
\bibitem [{\citenamefont {Martin}\ \emph {et~al.}(2014)\citenamefont {Martin},
  \citenamefont {Ishmukhametov}, \citenamefont {Hornung}, \citenamefont
  {Ahmad},\ and\ \citenamefont {Frasch}}]{Martin2014}%
  \BibitemOpen
  \bibfield  {author} {\bibinfo {author} {\bibfnamefont {J.~L.}\ \bibnamefont
  {Martin}}, \bibinfo {author} {\bibfnamefont {R.}~\bibnamefont
  {Ishmukhametov}}, \bibinfo {author} {\bibfnamefont {T.}~\bibnamefont
  {Hornung}}, \bibinfo {author} {\bibfnamefont {Z.}~\bibnamefont {Ahmad}}, \
  and\ \bibinfo {author} {\bibfnamefont {W.~D.}\ \bibnamefont {Frasch}},\
  }\href {\doibase 10.1073/pnas.1317784111} {\bibfield  {journal} {\bibinfo
  {journal} {Proc. Natl. Acad. Sci.}\ }\textbf {\bibinfo {volume} {111}},\
  \bibinfo {pages} {3715} (\bibinfo {year} {2014})}\BibitemShut {NoStop}%
\bibitem [{\citenamefont {Nguyen}\ \emph {et~al.}(2014)\citenamefont {Nguyen},
  \citenamefont {Klotsa}, \citenamefont {Engel},\ and\ \citenamefont
  {Glotzer}}]{Nguyen2014}%
  \BibitemOpen
  \bibfield  {author} {\bibinfo {author} {\bibfnamefont {N.~H.}\ \bibnamefont
  {Nguyen}}, \bibinfo {author} {\bibfnamefont {D.}~\bibnamefont {Klotsa}},
  \bibinfo {author} {\bibfnamefont {M.}~\bibnamefont {Engel}}, \ and\ \bibinfo
  {author} {\bibfnamefont {S.~C.}\ \bibnamefont {Glotzer}},\ }\href {\doibase
  10.1103/PhysRevLett.112.075701} {\bibfield  {journal} {\bibinfo  {journal}
  {Phys. Rev. Lett.}\ }\textbf {\bibinfo {volume} {112}},\ \bibinfo {pages}
  {075701} (\bibinfo {year} {2014})}\BibitemShut {NoStop}%
\bibitem [{\citenamefont {Yeo}\ \emph {et~al.}(2015)\citenamefont {Yeo},
  \citenamefont {Lushi},\ and\ \citenamefont {Vlahovska}}]{Yeo2015}%
  \BibitemOpen
  \bibfield  {author} {\bibinfo {author} {\bibfnamefont {K.}~\bibnamefont
  {Yeo}}, \bibinfo {author} {\bibfnamefont {E.}~\bibnamefont {Lushi}}, \ and\
  \bibinfo {author} {\bibfnamefont {P.~M.}\ \bibnamefont {Vlahovska}},\ }\href
  {\doibase 10.1103/PhysRevLett.114.188301} {\bibfield  {journal} {\bibinfo
  {journal} {Phys. Rev. Lett.}\ }\textbf {\bibinfo {volume} {114}},\ \bibinfo
  {pages} {188301} (\bibinfo {year} {2015})}\BibitemShut {NoStop}%
\bibitem [{\citenamefont {Soni}\ \emph {et~al.}(2018)\citenamefont {Soni},
  \citenamefont {Bililign}, \citenamefont {Magkiriadou}, \citenamefont
  {Sacanna}, \citenamefont {Bartolo}, \citenamefont {Shelley},\ and\
  \citenamefont {Irvine}}]{Soni2018}%
  \BibitemOpen
  \bibfield  {author} {\bibinfo {author} {\bibfnamefont {V.}~\bibnamefont
  {Soni}}, \bibinfo {author} {\bibfnamefont {E.}~\bibnamefont {Bililign}},
  \bibinfo {author} {\bibfnamefont {S.}~\bibnamefont {Magkiriadou}}, \bibinfo
  {author} {\bibfnamefont {S.}~\bibnamefont {Sacanna}}, \bibinfo {author}
  {\bibfnamefont {D.}~\bibnamefont {Bartolo}}, \bibinfo {author} {\bibfnamefont
  {M.~J.}\ \bibnamefont {Shelley}}, \ and\ \bibinfo {author} {\bibfnamefont
  {W.~T.~M.}\ \bibnamefont {Irvine}},\ }\href
  {https://arxiv.org/pdf/1812.09990.pdf http://arxiv.org/abs/1812.09990} {\
  (\bibinfo {year} {2018})},\ \Eprint {http://arxiv.org/abs/1812.09990}
  {arXiv:1812.09990} \BibitemShut {NoStop}%
\bibitem [{\citenamefont {Lenz}\ \emph {et~al.}(2004)\citenamefont {Lenz},
  \citenamefont {Joanny}, \citenamefont {J{\"{u}}licher},\ and\ \citenamefont
  {Prost}}]{Lenz2004}%
  \BibitemOpen
  \bibfield  {author} {\bibinfo {author} {\bibfnamefont {P.}~\bibnamefont
  {Lenz}}, \bibinfo {author} {\bibfnamefont {J.~F.}\ \bibnamefont {Joanny}},
  \bibinfo {author} {\bibfnamefont {F.}~\bibnamefont {J{\"{u}}licher}}, \ and\
  \bibinfo {author} {\bibfnamefont {J.}~\bibnamefont {Prost}},\ }\href
  {\doibase 10.1140/epje/i2003-10083-9} {\bibfield  {journal} {\bibinfo
  {journal} {Eur. Phys. J. E}\ }\textbf {\bibinfo {volume} {13}},\ \bibinfo
  {pages} {379} (\bibinfo {year} {2004})}\BibitemShut {NoStop}%
\bibitem [{\citenamefont {Jiko}\ \emph {et~al.}(2015)\citenamefont {Jiko},
  \citenamefont {Davies}, \citenamefont {Shinzawa-Itoh}, \citenamefont {Tani},
  \citenamefont {Maeda}, \citenamefont {Mills}, \citenamefont {Tsukihara},
  \citenamefont {Fujiyoshi}, \citenamefont {K{\"{u}}hlbrandt},\ and\
  \citenamefont {Gerle}}]{Jiko2015}%
  \BibitemOpen
  \bibfield  {author} {\bibinfo {author} {\bibfnamefont {C.}~\bibnamefont
  {Jiko}}, \bibinfo {author} {\bibfnamefont {K.~M.}\ \bibnamefont {Davies}},
  \bibinfo {author} {\bibfnamefont {K.}~\bibnamefont {Shinzawa-Itoh}}, \bibinfo
  {author} {\bibfnamefont {K.}~\bibnamefont {Tani}}, \bibinfo {author}
  {\bibfnamefont {S.}~\bibnamefont {Maeda}}, \bibinfo {author} {\bibfnamefont
  {D.~J.}\ \bibnamefont {Mills}}, \bibinfo {author} {\bibfnamefont
  {T.}~\bibnamefont {Tsukihara}}, \bibinfo {author} {\bibfnamefont
  {Y.}~\bibnamefont {Fujiyoshi}}, \bibinfo {author} {\bibfnamefont
  {W.}~\bibnamefont {K{\"{u}}hlbrandt}}, \ and\ \bibinfo {author}
  {\bibfnamefont {C.}~\bibnamefont {Gerle}},\ }\href {\doibase
  10.7554/eLife.06119} {\bibfield  {journal} {\bibinfo  {journal} {eLife}\ }
  (\bibinfo {year} {2015}),\ 10.7554/eLife.06119}\BibitemShut {NoStop}%
\bibitem [{\citenamefont {Lenz}\ \emph {et~al.}(2003)\citenamefont {Lenz},
  \citenamefont {Joanny}, \citenamefont {J{\"{u}}licher},\ and\ \citenamefont
  {Prost}}]{Lenz2003}%
  \BibitemOpen
  \bibfield  {author} {\bibinfo {author} {\bibfnamefont {P.}~\bibnamefont
  {Lenz}}, \bibinfo {author} {\bibfnamefont {J.-F.}\ \bibnamefont {Joanny}},
  \bibinfo {author} {\bibfnamefont {F.}~\bibnamefont {J{\"{u}}licher}}, \ and\
  \bibinfo {author} {\bibfnamefont {J.}~\bibnamefont {Prost}},\ }\href
  {\doibase 10.1103/PhysRevLett.91.108104} {\bibfield  {journal} {\bibinfo
  {journal} {Phys. Rev. Lett.}\ }\textbf {\bibinfo {volume} {91}},\ \bibinfo
  {pages} {108104} (\bibinfo {year} {2003})}\BibitemShut {NoStop}%
\bibitem [{\citenamefont {Chaikin}\ and\ \citenamefont
  {Lubensky}(1995)}]{Chaikin}%
  \BibitemOpen
  \bibfield  {author} {\bibinfo {author} {\bibfnamefont {P.~M.}\ \bibnamefont
  {Chaikin}}\ and\ \bibinfo {author} {\bibfnamefont {T.~C.}\ \bibnamefont
  {Lubensky}},\ }\href@noop {} {\emph {\bibinfo {title} {{Principles of
  condensed matter physics}}}}\ (\bibinfo  {publisher} {Cambridge University
  Press},\ \bibinfo {year} {1995})\BibitemShut {NoStop}%
\bibitem [{\citenamefont {Lushi}\ and\ \citenamefont
  {Vlahovska}(2015)}]{Lushi2015}%
  \BibitemOpen
  \bibfield  {author} {\bibinfo {author} {\bibfnamefont {E.}~\bibnamefont
  {Lushi}}\ and\ \bibinfo {author} {\bibfnamefont {P.~M.}\ \bibnamefont
  {Vlahovska}},\ }\href {\doibase 10.1007/s00332-015-9254-9} {\bibfield
  {journal} {\bibinfo  {journal} {J. Nonlinear Sci.}\ }\textbf {\bibinfo
  {volume} {25}},\ \bibinfo {pages} {1111} (\bibinfo {year}
  {2015})}\BibitemShut {NoStop}%
\bibitem [{\citenamefont {Noether}(1918)}]{Noether1918}%
  \BibitemOpen
  \bibfield  {author} {\bibinfo {author} {\bibfnamefont {E.}~\bibnamefont
  {Noether}},\ }\href@noop {} {\bibfield  {journal} {\bibinfo  {journal} {Nachr
  D K{\"{o}}nig Gesellsch D Wiss Zu G{\"{o}}ttingen Mathphys Klasse}\ ,\
  \bibinfo {pages} {235}} (\bibinfo {year} {1918})},\ \Eprint
  {http://arxiv.org/abs/0503066} {0503066 [physics]} \BibitemShut {NoStop}%
\bibitem [{\citenamefont {Falkovich}(2009)}]{Falkovich2009}%
  \BibitemOpen
  \bibfield  {author} {\bibinfo {author} {\bibfnamefont {G.}~\bibnamefont
  {Falkovich}},\ }\href {\doibase 10.1088/1751-8113/42/12/123001} {\bibfield
  {journal} {\bibinfo  {journal} {J. Phys. A:}\ }\textbf {\bibinfo {volume}
  {42}},\ \bibinfo {pages} {123001} (\bibinfo {year} {2009})}\BibitemShut
  {NoStop}%
\bibitem [{\citenamefont {Conti}\ and\ \citenamefont
  {Badin}(2017)}]{Conti2017}%
  \BibitemOpen
  \bibfield  {author} {\bibinfo {author} {\bibfnamefont {G.}~\bibnamefont
  {Conti}}\ and\ \bibinfo {author} {\bibfnamefont {G.}~\bibnamefont {Badin}},\
  }\href {\doibase 10.3390/fluids2040050} {\bibfield  {journal} {\bibinfo
  {journal} {Fluids}\ }\textbf {\bibinfo {volume} {2}},\ \bibinfo {pages} {50}
  (\bibinfo {year} {2017})}\BibitemShut {NoStop}%
\bibitem [{\citenamefont {Aref}\ and\ \citenamefont
  {Pomphrey}(1982)}]{Aref1982}%
  \BibitemOpen
  \bibfield  {author} {\bibinfo {author} {\bibfnamefont {H.}~\bibnamefont
  {Aref}}\ and\ \bibinfo {author} {\bibfnamefont {N.}~\bibnamefont
  {Pomphrey}},\ }\href {\doibase 10.1098/rspa.1982.0047} {\bibfield  {journal}
  {\bibinfo  {journal} {Proc. Roy. Soc. A}\ }\textbf {\bibinfo {volume}
  {380}},\ \bibinfo {pages} {359} (\bibinfo {year} {1982})}\BibitemShut
  {NoStop}%
\bibitem [{\citenamefont {Wigner}(1934)}]{Wigner1934}%
  \BibitemOpen
  \bibfield  {author} {\bibinfo {author} {\bibfnamefont {E.}~\bibnamefont
  {Wigner}},\ }\href {\doibase 10.1103/PhysRev.46.1002} {\bibfield  {journal}
  {\bibinfo  {journal} {Physical Review}\ }\textbf {\bibinfo {volume} {46}},\
  \bibinfo {pages} {1002} (\bibinfo {year} {1934})}\BibitemShut {NoStop}%
\bibitem [{\citenamefont {Nelson}(2002)}]{Nelson2002}%
  \BibitemOpen
  \bibfield  {author} {\bibinfo {author} {\bibfnamefont {D.~R.}\ \bibnamefont
  {Nelson}},\ }\href@noop {} {\emph {\bibinfo {title} {{Defects and geometry in
  condensed matter physics}}}}\ (\bibinfo  {publisher} {Cambridge University
  Press},\ \bibinfo {year} {2002})\BibitemShut {NoStop}%
\bibitem [{\citenamefont {Delaunay}(1934)}]{Delaunay1934}%
  \BibitemOpen
  \bibfield  {author} {\bibinfo {author} {\bibfnamefont {P.~B.}\ \bibnamefont
  {Delaunay}},\ }\href@noop {} {\bibfield  {journal} {\bibinfo  {journal}
  {Bull. Acad. Sci. USSR,}\ ,\ \bibinfo {pages} {793}} (\bibinfo {year}
  {1934})}\BibitemShut {NoStop}%
\bibitem [{\citenamefont {Sokolov}\ and\ \citenamefont
  {Diamant}(2018)}]{Sokolov2018}%
  \BibitemOpen
  \bibfield  {author} {\bibinfo {author} {\bibfnamefont {Y.}~\bibnamefont
  {Sokolov}}\ and\ \bibinfo {author} {\bibfnamefont {H.}~\bibnamefont
  {Diamant}},\ }\href@noop {} {\bibfield  {journal} {\bibinfo  {journal} {J.
  Chem. Phys.}\ }\textbf {\bibinfo {volume} {149}},\ \bibinfo {pages} {034901}
  (\bibinfo {year} {2018})},\ \Eprint {http://arxiv.org/abs/1804.08092}
  {arXiv:1804.08092} \BibitemShut {NoStop}%
\bibitem [{\citenamefont {Kinosita}\ \emph {et~al.}(1998)\citenamefont
  {Kinosita}, \citenamefont {Yasuda}, \citenamefont {Noji}, \citenamefont
  {Ishiwata},\ and\ \citenamefont {Yoshida}}]{Kinosita1998}%
  \BibitemOpen
  \bibfield  {author} {\bibinfo {author} {\bibfnamefont {K.}~\bibnamefont
  {Kinosita}}, \bibinfo {author} {\bibfnamefont {R.}~\bibnamefont {Yasuda}},
  \bibinfo {author} {\bibfnamefont {H.}~\bibnamefont {Noji}}, \bibinfo {author}
  {\bibfnamefont {S.}~\bibnamefont {Ishiwata}}, \ and\ \bibinfo {author}
  {\bibfnamefont {M.}~\bibnamefont {Yoshida}},\ }\href {\doibase
  10.1016/S0092-8674(00)81142-3} {\bibfield  {journal} {\bibinfo  {journal}
  {Cell}\ }\textbf {\bibinfo {volume} {93}},\ \bibinfo {pages} {21} (\bibinfo
  {year} {1998})}\BibitemShut {NoStop}%
\bibitem [{\citenamefont {Levitt}\ \emph {et~al.}(2009)\citenamefont {Levitt},
  \citenamefont {Kuimova}, \citenamefont {Yahioglu}, \citenamefont {Chung},
  \citenamefont {Suhling},\ and\ \citenamefont {Phillips}}]{Levitt2009}%
  \BibitemOpen
  \bibfield  {author} {\bibinfo {author} {\bibfnamefont {J.~A.}\ \bibnamefont
  {Levitt}}, \bibinfo {author} {\bibfnamefont {M.~K.}\ \bibnamefont {Kuimova}},
  \bibinfo {author} {\bibfnamefont {G.}~\bibnamefont {Yahioglu}}, \bibinfo
  {author} {\bibfnamefont {P.-H.}\ \bibnamefont {Chung}}, \bibinfo {author}
  {\bibfnamefont {K.}~\bibnamefont {Suhling}}, \ and\ \bibinfo {author}
  {\bibfnamefont {D.}~\bibnamefont {Phillips}},\ }\href {\doibase
  10.1021/jp9013493} {\bibfield  {journal} {\bibinfo  {journal} {J. Phys. Chem.
  C}\ }\textbf {\bibinfo {volume} {113}},\ \bibinfo {pages} {11634} (\bibinfo
  {year} {2009})}\BibitemShut {NoStop}%
\bibitem [{\citenamefont {Blum}\ \emph {et~al.}(2019)\citenamefont {Blum},
  \citenamefont {Hahn}, \citenamefont {Meier}, \citenamefont {Davies},\ and\
  \citenamefont {K{\"{u}}hlbrandt}}]{Blum2019}%
  \BibitemOpen
  \bibfield  {author} {\bibinfo {author} {\bibfnamefont {T.~B.}\ \bibnamefont
  {Blum}}, \bibinfo {author} {\bibfnamefont {A.}~\bibnamefont {Hahn}}, \bibinfo
  {author} {\bibfnamefont {T.}~\bibnamefont {Meier}}, \bibinfo {author}
  {\bibfnamefont {K.~M.}\ \bibnamefont {Davies}}, \ and\ \bibinfo {author}
  {\bibfnamefont {W.}~\bibnamefont {K{\"{u}}hlbrandt}},\ }\href {\doibase
  10.1073/pnas.1816556116} {\bibfield  {journal} {\bibinfo  {journal} {Poc.
  Natl. Acad. Sci.}\ ,\ \bibinfo {pages} {201816556}} (\bibinfo {year}
  {2019})}\BibitemShut {NoStop}%
\bibitem [{\citenamefont {Almendro-Vedia}\ \emph {et~al.}(2017)\citenamefont
  {Almendro-Vedia}, \citenamefont {Natale}, \citenamefont {Mell}, \citenamefont
  {Bonneau}, \citenamefont {Monroy}, \citenamefont {Joubert},\ and\
  \citenamefont {L{\'{o}}pez-Montero}}]{Almendro-Vedia2017}%
  \BibitemOpen
  \bibfield  {author} {\bibinfo {author} {\bibfnamefont {V.~G.}\ \bibnamefont
  {Almendro-Vedia}}, \bibinfo {author} {\bibfnamefont {P.}~\bibnamefont
  {Natale}}, \bibinfo {author} {\bibfnamefont {M.}~\bibnamefont {Mell}},
  \bibinfo {author} {\bibfnamefont {S.}~\bibnamefont {Bonneau}}, \bibinfo
  {author} {\bibfnamefont {F.}~\bibnamefont {Monroy}}, \bibinfo {author}
  {\bibfnamefont {F.}~\bibnamefont {Joubert}}, \ and\ \bibinfo {author}
  {\bibfnamefont {I.}~\bibnamefont {L{\'{o}}pez-Montero}},\ }\href {\doibase
  10.1073/pnas.1701207114} {\bibfield  {journal} {\bibinfo  {journal} {Proc.
  Natl. Acad. Sci.}\ }\textbf {\bibinfo {volume} {114}},\ \bibinfo {pages}
  {11291} (\bibinfo {year} {2017})}\BibitemShut {NoStop}%
\end{thebibliography}%
\bibliographystyle{apsrev4-1}

\end{document}